\documentclass[aps,pra,letterpaper,10pt,twocolumn,showpacs,superscriptaddress]{revtex4-1}

\usepackage{amsfonts,amssymb,amsmath,amsthm,graphicx}
\usepackage{url}
\usepackage{bbm}
\usepackage[usenames,dvipsnames]{xcolor}
\usepackage{nicefrac}
\usepackage{multirow}
\usepackage{natbib}
\usepackage{dblfloatfix} 
\usepackage{setspace}
\usepackage{subfigure}
\usepackage{comment}
\usepackage[colorlinks=true,bookmarks=false,citecolor=blue,urlcolor=blue]{hyperref}
\usepackage[normalem]{ulem} 
\usepackage{todonotes}

\newcommand{\beq}{\begin{equation}}
\newcommand{\eeq}{\end{equation}}
\newcommand{\Sclock}{\mathcal{S}_{\rm clk}(t)}
\newcommand{\Sdet}{\mathcal{S}_{\rm det}(t)}
\newcommand{\tauclk}{\tau_{\rm clk}}

\newcommand{\red}{\color{red}}

\def\virgolette #1{``#1"}

\newcommand{\soutP}[1]{{\red\sout{#1}}}
\renewcommand{\soutP}[1]{{}}

\begin{document}
\raggedbottom
\title{Efficient random number generation techniques for CMOS SPAD array based devices}

\author{Andrea Stanco}
\affiliation{Dipartimento di Ingegneria dell'Informazione, Universit\`a degli Studi di Padova, via Gradenigo 6B, 35131 Padova, Italia}

\author{Davide G.~Marangon}
\altaffiliation{Present address: Toshiba Research Europe Ltd, Cambridge Research Laboratory, 208 Cambridge Science Park, Milton Road, Cambridge, CB4 0GZ, UK}
\affiliation{Dipartimento di Ingegneria dell'Informazione, Universit\`a degli Studi di Padova, via Gradenigo 6B, 35131 Padova, Italia}

\author{Giuseppe Vallone}
\affiliation{Dipartimento di Ingegneria dell'Informazione, Universit\`a degli Studi di Padova, via Gradenigo 6B, 35131 Padova, Italia}
\affiliation{Dipartimento di Fisica e Astronomia, Universit\`a degli Studi di Padova, via Marzolo 8, 35131 Padova, Italy}

\author{Samuel Burri}
\affiliation{EPFL, Route Cantonale, 1015 Lausanne, Switzerland}

\author{Edoardo Charbon}
\affiliation{EPFL, Route Cantonale, 1015 Lausanne, Switzerland}

\author{Paolo Villoresi}
\email{paolo.villoresi@dei.unipd.it}
\affiliation{Dipartimento di Ingegneria dell'Informazione, Universit\`a degli Studi di Padova, via Gradenigo 6B, 35131 Padova, Italia}

\date{\today}


\begin{abstract}
This work presents new techniques to produce true random bits by exploiting single photon time of arrival. Two FPGA-based QRNG devices are presented: Randy which uses one discrete SPAD and LinoSPAD which uses a CMOS SPAD array, along with a time-to-digital converter (TDC). Post-processing procedures are explained in order to extract randomness taking care of SPAD and TDC non-idealities. These procedures are based on the application of Peres~\cite{Peres1992} and
Zhou-Bruk~\cite{Zhou2012}
 extraction algorithms. Achieved generation rates are 1.8 Mbit/s for Randy device and 310 Mbit/s for LinoSPAD device. Randy QRNG also features a real time procedure which was used for the realization of fundamental tests of physics.

\end{abstract}

\maketitle

\section{Introduction}
In very recent years, there has been a widespread interest in random number generators based on physical processes of quantum nature.
In fact, these devices, the so-called quantum random number generators (QRNGs), represent the ultimate way to obtain reliable randomness, free from the typical non-random issue that affects algorithmic random number generators. 
Typically, QRNGs exploit the quantum properties of the optical radiation field in many \virgolette{recipes}.
The different architectures exploit different properties; setups using entangled systems and violating Bell inequalities feature the highest unpredictability in the so-called device-independent (DI) framework \cite{Pironio2010,Christensen2013,Bierhorst2018,Liu2018,Gomez2018}.
Systems that trust the measurement apparatus but not the states of the quantum system or vice versa, the so-called semi-device-independent generators, work under less strict assumptions \cite{Ma2016}: for this reason, they feature, in principle, less unpredictability than DI-QRNG but are more feasible to implement and reach even larger generation rates.
The last category includes those generators that work in a framework of complete trust in both the quantum system and the measurement apparatus.
This means that the absence of side information, exploitable by an adversary, is assumed.
The most famous example of this kind of generator is the \emph{welcher weg} QRNG that produces randomness according to which path a photon takes after interacting with a beam-splitter \cite{Rarity1994,Jennewin2000}.
In this work we consider the sub-category of trusted QRNG that were developed in order to limit the number of single photon detectors.
Random number generation with just one detector is indeed possible by exploiting time as an additional degree of freedom and by leveraging on the statistical features of the photon detection distribution. 
This could be done with the following procedure: time sampling of the photon detector with a sampling rate almost equal to the photon rate; application of dedicated generation protocols; application of dedicated unbiasing algorithms.
In this work we will show that it is possible to generate true random numbers without the application of dedicated generation protocols.  Using a higher sampling rate we applied unbiasing algorithms directly to the samples stream achieving higher generation rate of true random numbers. We used this technique on two different systems: the first one, Randy, uses a standard clock to sample the signal of one single photon detector; the second one, LinoSPAD, uses both a standard clock and a time-to-digital converter (TDC) to sample the signals from a matrix of 256 single photon detectors.

The paper is structured as follows: in Sec. \ref{examples} the most common ways to generate random numbers from time will be reviewed. 
In Sec. \ref{novel} we will introduce our method and we will compare it with the previous ones. 
The results achieved by applying it to the numbers obtained with a single detector connected to an FPGA will be presented.
In Sec. \ref{linospadSec} we will extend this method to a matrix of 256 detectors with time tagging capabilities.
In Sec. \ref{conclusions} the conclusions will be presented.

\section{Existing paradigms of single detector QRNGs}\label{examples}
We now introduce two elaborated generation protocols, reported in literature, that can extract randomness from a photons distribution.

As a first example, let us consider the generation protocol introduced by Stip\v{c}evi\'{c} \emph{et al.} in \cite{Stipc2007}, denominated here \virgolette{Diff-QRNG}. 
The Diff-QRNG comprises: a light source attenuated to single photon level illuminating a single photon detector; a clock that counts the time between the detection events. 
A binary random variable $X_j=\left\{0,1\right\}$ is obtained by comparing the length of time intervals between three consecutive detections. 
It is therefore convenient to define the discrete random variable $\mathcal{T}^j$, associated to the detection instants, such that $\mathcal{T}^j=\left\{t^j_{0},t^j_{1},t^j_{2}\right\}$.
Hence,  $X_j$ takes its value $x_j$ according to the following \virgolette{rule}: given $\Delta T_1 = t_{1}-t_{0}$ and  $\Delta T_2 = t_{2}-t_{1}$,
\begin{itemize}
\item if $\Delta T_1>\Delta T_2$ then $x_j=0$ 
\item if $\Delta T_1<\Delta T_2$ then $x_j=1$ 
\item if $\Delta T_1=\Delta T_2$ then $x_j=\emptyset$, i.e. no bit is generated.
\end{itemize}
The rule is iterated so that for the next bit $b_{j+1}$ the new time interval starts with the end of the previous one, i.e., $t^j_{2}=t^{j+1}_{0}$.
The physical principle that guarantees the identical ($\text{Pr}\left[x_j=0\right] = \text{Pr}\left[x_j=1\right]$) and independent ($\text{Pr}\left[x_j|x_{j-1}\right] = \text{Pr}\left[x_j|x_{j-1},x_{j-2},\dots,x_1\right]$=\nicefrac{1}{2}) distribution of the bits follows from the memoryless property that characterizes the exponential distribution of the interarrival times $\Delta T$, namely $\text{Pr}\left[\Delta T_1>\Delta T_2\right] = \text{Pr}\left[\Delta T_2>\Delta T_1\right]=\nicefrac{1}{2}$.
As a consequence, a random string $X=\left\{X_1, X_2,\dots, X_n\right\}$ with $n\rightarrow \infty$, is characterized by full Shannon entropy, i.e. $H(X)=1$ bit.
This implies that the average number of i.i.d. bits that are generated per unit time, is equal to $r_\text{i.i.d.}=r_{\text{phot}}/2$, where $r_{\text{phot}}$ is the number of photo-detections per unit time \footnote{With the exception of bit $x_1$ and $x_n$ two photo-detection are necessary to generate an i.i.d. bit.}.

As a second example, let us consider the generation protocol introduced by F\"{u}rst \emph{et al.} in \cite{Furst2010}, denominated here \virgolette{OdEven-QRNG}. 
As in the previous example, in the OdEven-QRNG a light source attenuated to single photon level illuminates a photomultiplier but in this case a counter enumerates the number of photons detected within a fixed time interval, $\tau$, corresponding to the period of a sampling signal.
Defining $n_\tau^j$ as the number of detections within the interval $\tau_j$, a random binary variable $X_j$ assumes its value $x_j$, with the following rule: 
\begin{itemize}
\item if $n_\tau^j \mod 2 = 0$ then $x_j=0$
\item if $n_\tau^j \mod 2 = 1$ then $x_j=1$,
\end{itemize}
in other words, the bit value is determined according whether an even or odd number of detections is registered in the time interval $\tau$. For a Poisson distribution we can write the probability of having an even or odd number of detections:
\begin{itemize}
\item $\text{Pr}[2n]=\sum_{n=0}^{\infty}\frac{(\lambda\tau)^{2n}}{(2n)!}e^{-\lambda\tau}=\frac{1+e^{-2\lambda\tau}}{2}=\text{Pr}[x_j=0]$
\item $\text{Pr}[2n+1]=\sum_{n=0}^{\infty}\frac{(\lambda\tau)^{2n+1}}{(2n+1)!}e^{-\lambda\tau}=\frac{1-e^{-2\lambda\tau}}{2}=\text{Pr}[x_j=1]$
\end{itemize}
where $\lambda$ is the mean number of photons per second and $\lambda\tau=\langle n_\tau \rangle$ is the mean number of photons per time interval $\tau$. To avoid a bias in the output, $\langle n_\tau \rangle$ has to be sufficiently large so that $\text{Pr}[x_j=0] \simeq \text{Pr}[x_j=1]$.
Therefore, since $H(X)$ is a function of $\langle n_\tau \rangle$, the generation rate is given by $R=H(X)/\tau$.

\begin{figure}[htbp]
\includegraphics[width=\columnwidth]{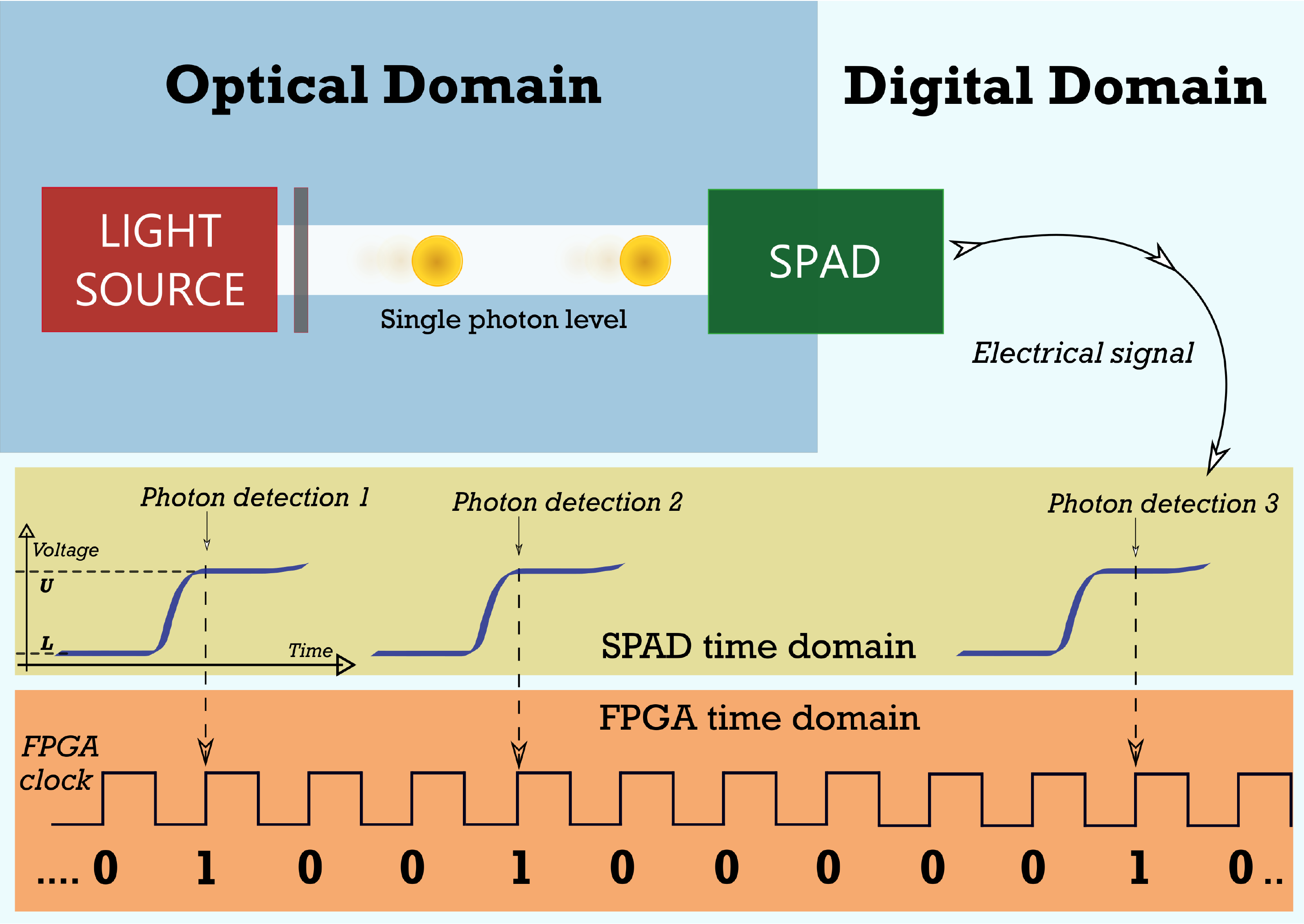}
\caption{A schematic view of Randy system. The SPAD links the optical domain to the digital domain by detecting photons from a light source \- attenuated to single photon level \- and by sending electrical signals to the FPGA. Through its internal clock, the FPGA samples the SPAD events by its own time domain. Clock and SPAD rates are not on scale.}
\label{schematic}
\end{figure}

\section{The novel approach}\label{novel}

Our contribution takes into consideration the simplest and most efficient way to generate random numbers by using the temporal degree of freedom without the need to devise complex \virgolette{rules}.
In its essence, the process of random number generation with a single-photon detector, for instance a single-photon avalanche diode (SPAD), can be considered as a process with two signals: 
a squared-wave signal $\Sdet$ generated by the single-photon detector that is sampled by a periodic signal $\Sclock$, which is generated by a clock with a period $\tauclk$.
Without impinging photons $\Sdet$ has a typical value $L$.
Given a photo-detection at the time instant $t_i$, the $\Sdet$ toggles its state from  $\Sdet(t<t_i)=L$ to $\Sdet(t_i \leq t < \tau_U )=U$. $\tau_U$ is the specific fixed time interval in which a SPAD keeps the state $U$ before returning to $L$ and typically $\tau_U <\tau_{\text{dead}}$, being $\tau_{\text{dead}}$ the SPAD dead-time. This means that $\Sdet$ toggles back to the $L$ state before the SPAD being able to detect another photon.
The binary random variable $X_j$ takes its value $x_j$ at the instant $j\tauclk$ according to the following rule: 
\begin{equation}
x_j=\begin{cases}
1 & {\rm if\ }\mathcal{S}_{\text{det}}(j\, \tauclk )=U\neq
\mathcal{S}_{\text{det}}((j-1)\, \tauclk )
\\
0 & {\rm otherwise}
\end{cases}
\end{equation}
with the index $j \in\left\{0,1,2,\dots\right\}$.
The above rule is equivalent to having $x_j=1$ whenever the clock detects a rising edge of $\Sdet$ and
$x_j=0$ otherwise.
It is worth to be noticed that all existing paradigms of QRNG based on
photon time of arrival can be seen as a (non-optimal) post-processing
algorithm of the sequence $X=(x_1,x_2,\cdots,x_n,\cdots)$. We note that the maximum content of randomness that can be extracted by the 
above physical process is fully included in the $X$ sequence, and in particular it is given by the Shannon entropy (in the large $n$ limits):
\beq 
H(X)=-p_0\log_{2}(p_0)-p_1\log_{2}(p_1)
\eeq 
where $p_0$ and $p_1$ are the probability of obtaining $X=0$ and $X=1$ respectively.

It is clear that the maximum generation rate is given by $r_{\rm max}=\tauclk^{-1}$. Given a mean photon number per second $\lambda =r_{\text{phot}}$, for a Poisson distribution the probability of no detections within the $\tauclk$ interval is equal to $\text{Pr}[\emptyset]=e^{-r_{\text{phot}}\tauclk}$. 
To achieve the rate $r_{\rm max}$ it is necessary to avoid a bias in the output string $X$, by tuning
the photon detection rate in order to obtain at least one detection within a sampling period, namely $1-\text{Pr}[\emptyset]=1/2$ from which we obtain $r_{\text{phot}}=\ln{2} / \tauclk$. 
It is therefore clear that the faster the clock rate is, the larger the detection rate should be to keep low the bias in 0.
However, $r_{\text{phot}}$ cannot be set arbitrarily large as it is strongly limited by the detector dead time.  
Given a fixed $r_{\rm phot}$, our idea is to increase the generation rate $r$ by deliberately increasing the sampling rate and producing a highly biased string with plenty of 0's and very few 1's. Then, we re-balance the bias through an optimal post-processing algorithm, introduced by Peres \cite{Peres1992}. The Peres algorithm is a revision of the famous von Neumann algorithm \cite{Neumann1951} and allows to make a biased string uniform while distilling the maximum entropy from that. We give a brief description of it.
Given a sequence of samples $s_{w}$, the Peres procedure is defined as 
\begin{equation*}
\Psi(s_{w})=\Psi_{N}(s_{w})||\Psi(\Psi_{U}(s_{w}))||\Psi(\Psi_{V}(s_{w}))
\end{equation*}
where $\Psi_{N}(s)$ is the von Neumann algorithm applied to the string, $\Psi(\Psi_{U}(s))$ is the Peres algorithm applied to a new sub-string $\Psi_{U}(s)$ and $\Psi(\Psi_{V}(s))$ is the Peres algorithm applied to another new sub-string $\Psi_{V}(s)$. $\Psi_{U}(s)$ is created by an XOR operation on samples pair while $\Psi_{V}(s)$ is created by taking the latest samples of every 00 and 11 pairs. For a full description of the procedure refer to \cite{Peres1992}.
Clearly, the i.i.d. property comes from the Poisson distribution itself.

\begin{figure}[t]
\includegraphics[width=\columnwidth]{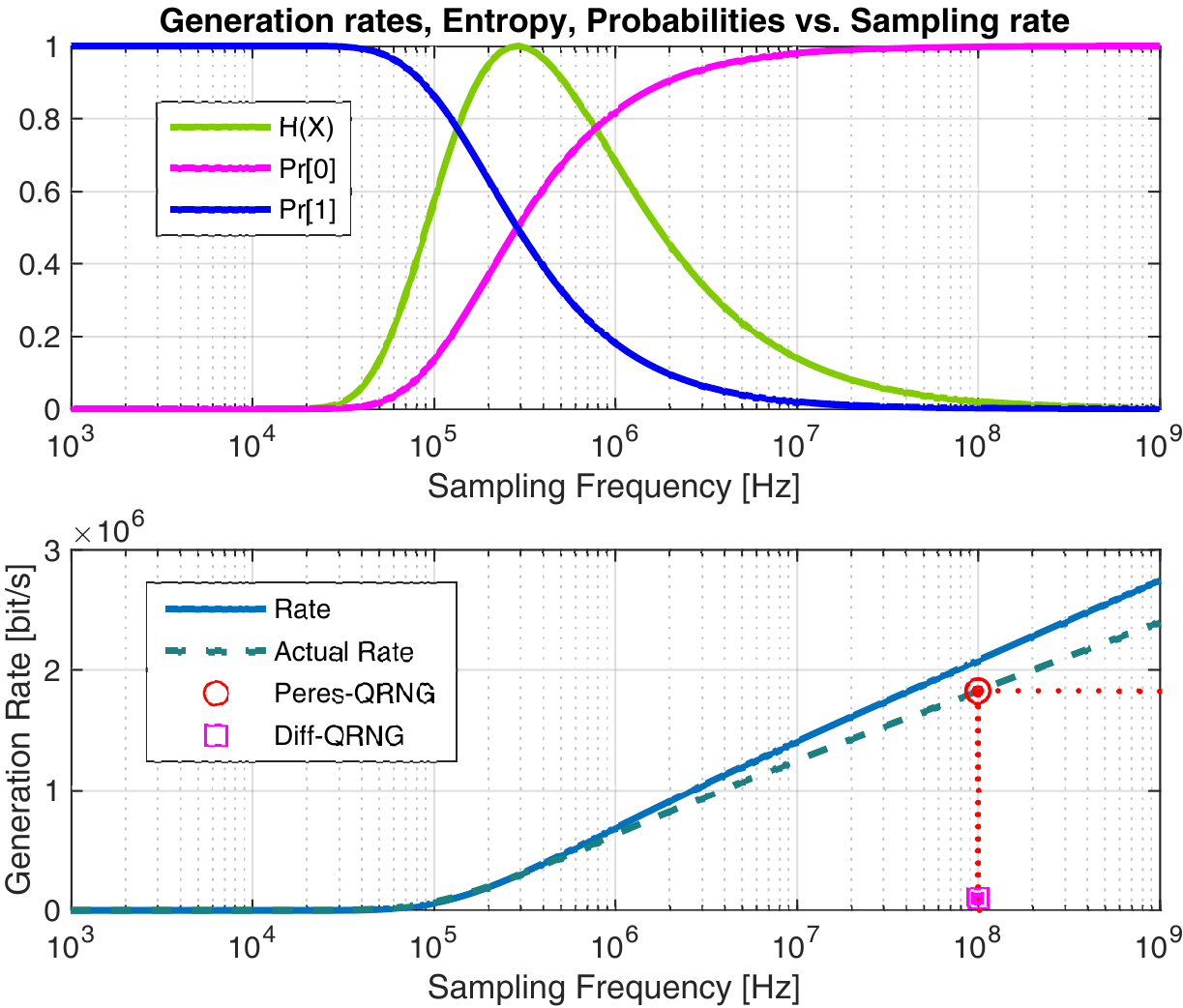}
\caption{The upper plot shows the balance between 0's and 1's ($\text{Pr}[\emptyset]$ and $\text{Pr}[1]$) as well as the binary entropy as function of the sampling rate. The photon count rate is fixed to 200 kcounts/s. In the lower plot the continuous line shows the theoretical generation rate as function of the sampling rate before any afterpulses or dead time treatment.  The dashed line shows the actual generation after the afterpulses and dead time removal. The expected 1.815 Mbit/s rate after Peres post-processing is highlighted by the dotted line. The magenta square highlights the rate obtained by Diff-QRNG protocol.}
\label{ratePeres}
\end{figure}

The extraction rate after Peres algorithm, under asymptotic assumptions, is given by 
\beq 
r_{\text{i.i.d.}} = \tauclk^{-1}\times H(X)\,
\eeq
As shown in Fig. \ref{ratePeres}, by fixing $r_{\rm phot}$, the extraction rate increases with the sampling rate despite the value of $H(X)$ (the maximum entropy is reached with a sampling rate equal to $r_{\text{phot}}$/ln2 = 288.539 kHz). These plots confirm the following: with a fixed $r_{\text{phot}}$, the generation rate is more influenced by the actual input length than by the bias value. Therefore, the optimal choice is to have $\tauclk^{-1}>>r_{\text{phot}}$.

To summarize, any QRNG based on the photon time of arrival can be modelled by a binary signal $\Sdet$ sampled by a clock $\Sclock$ that gives an output sequence $X$. Standard generation protocols, such as Diff-QRNG or OdEven-QRNG, can be applied to $X$. Nevertheless, these protocols are far from be efficient. We here propose an optimal post-processing able to extract the maximum available entropy from the string $X$ based on the Peres algorithm. In the next section we will show how to apply our method to physical generators, while taking care of the non-idealities of the detectors.

\section{Single detector implementation}
We first implemented our method by using a single SPAD shined by an attenuated laser light. We designed a system that is capable to produce true random numbers using different generation protocols as well as Peres algorithm. The system was developed from an FPGA/CPU device \footnote{We used the ZedBoard produced by Avnet.}. The FPGA allows us a full control over the generation process and a great flexibility in order to easily switch from one protocol to another. We called this design \textit{RANDY}. A schematic view of the setup is shown in Fig. \ref{schematic}. We set the light source intensity to keep the photon count rate around 200 kcounts/s, due to the SPAD non-linear behaviour on higher rates. As a first implementation we used the default 100 MHz system clock to sample the SPAD signals. The advantages are quite clear since an FPGA allows a full description of the time evolution of the system: every operation can be described as a multiple of the fundamental time unit $\tauclk$ defined by the system clock. Therefore, the behaviour of the system is fully deterministic apart from the non-deterministic side due to the true randomness of the photon time of arrival. 

The FPGA saves a logical 0 for every clock cycle whenever the SPAD signal is low, a logical 1 otherwise.
The raw string is then post-processed on a dedicated CPU. Due to the huge difference between the clock rate and the photon rate (100 MHz over 200 kcount/s), the raw strings are heavily biased in zero with a percentage of 99.8\%. Ideally, the role of Peres algorithm is to reduce the bias to zero. Nevertheless, Peres algorithm cannot work around any correlation. On the contrary, it could emphasize it. Therefore, any correlation has to be removed before the application of Peres algorithm. In our case, correlation comes from afterpulses and dead time of the detectors and we describe how to remove them in the following.

\begin{figure}[htbp]
\centering
\centerline{\includegraphics[width=\columnwidth]{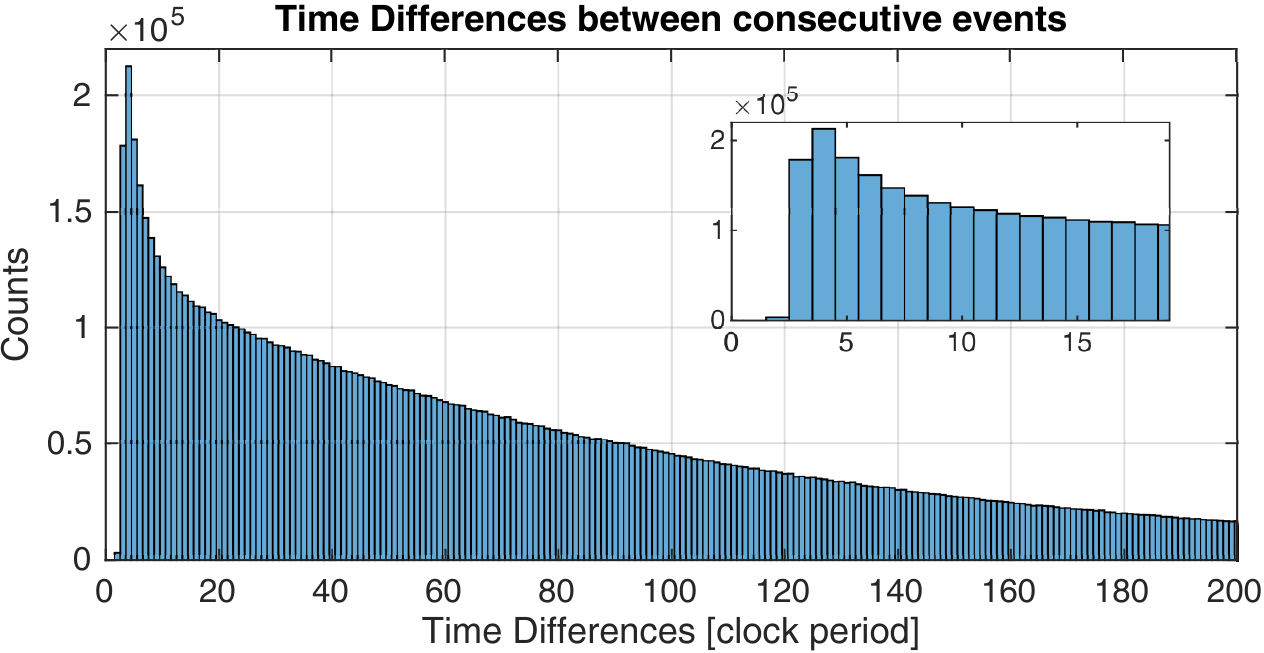}}
\centerline{\includegraphics[width=\columnwidth]{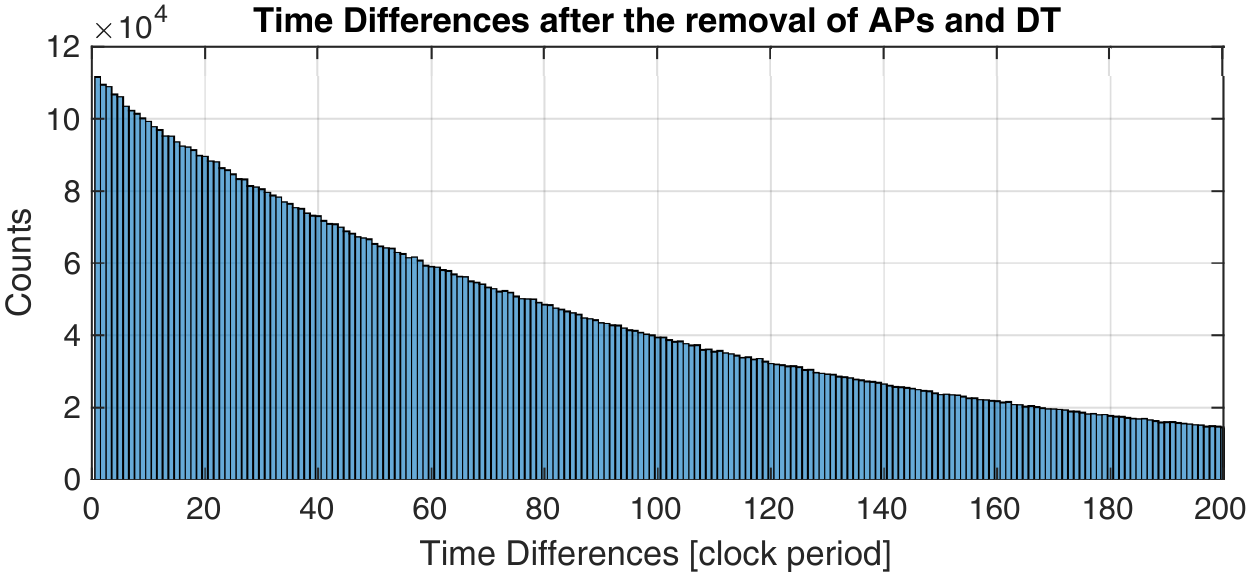}}
\caption{Top plot: time differences histogram before any post-processing. It is clear the presence of a dead time since there are no events that are less than 3 clock cycles far. It is also clear the presence of afterpulses. The curve trend shows a higher count rate below the 18 clock cycles threshold. Bottom Plot: time difference histogram after the removal of afterpulses and dead time.}
\label{diffRandy} 
\end{figure}

We evaluated the distribution of the time interval between consecutive events. For ideal Poissonian events, this distribution is expected to be a decreasing exponential. As shown in Fig. \ref{diffRandy}, the experimental distribution differs from the ideal distribution by two effects.

The first is the {\it dead time}, whose value depends on the specific SPAD used \footnote{We used the SPCM-ARQH by Excelitas.} and represents the minimum time distance between two rising edge of $\Sdet$. $\tau_{\text{dead}}\simeq 30\,{\rm ns}$ corresponding to 3 clock cycles. 

The second effect is the {\it afterpulse}. Afterpulses are spurious events that occur randomly within a fixed time interval from a real detection. As a result, they produce a peak which undermines the exponential trend of the events difference distribution. We evaluated a 18 clock cycles (180 ns) cross-point between the only-true-events region and the afterpulses-region. 

Dead-time and afterpulses introduce correlations in the output string. Clearly, after a value $x_j=1$, some of the few following bits in $X$ are not independent. 
Since we estimated 18 clock cycles as the required time to be in the true-event region, to remove the correlation we may simply remove 18 values of $X$ following any value $x_j=1$.

\begin{figure}[htbp]
\centering
\centerline{\includegraphics[width=\columnwidth]{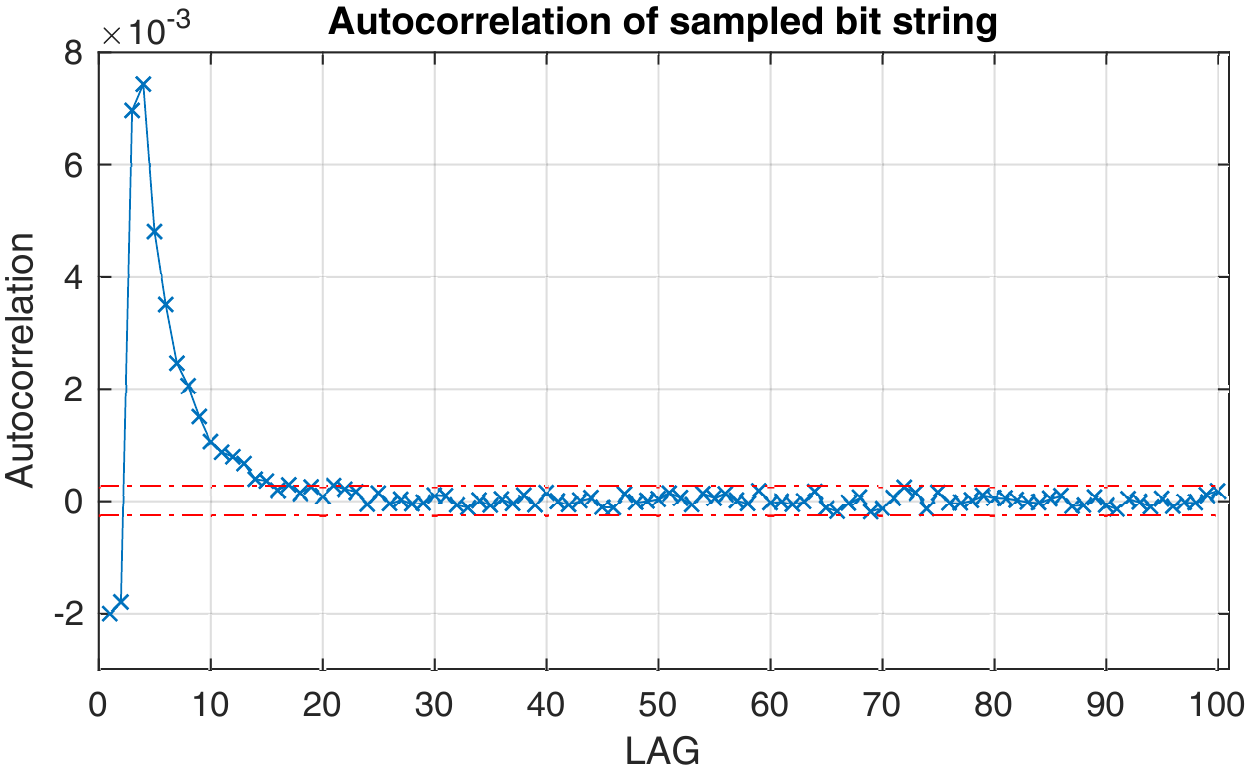}}
\centerline{\includegraphics[width=\columnwidth]{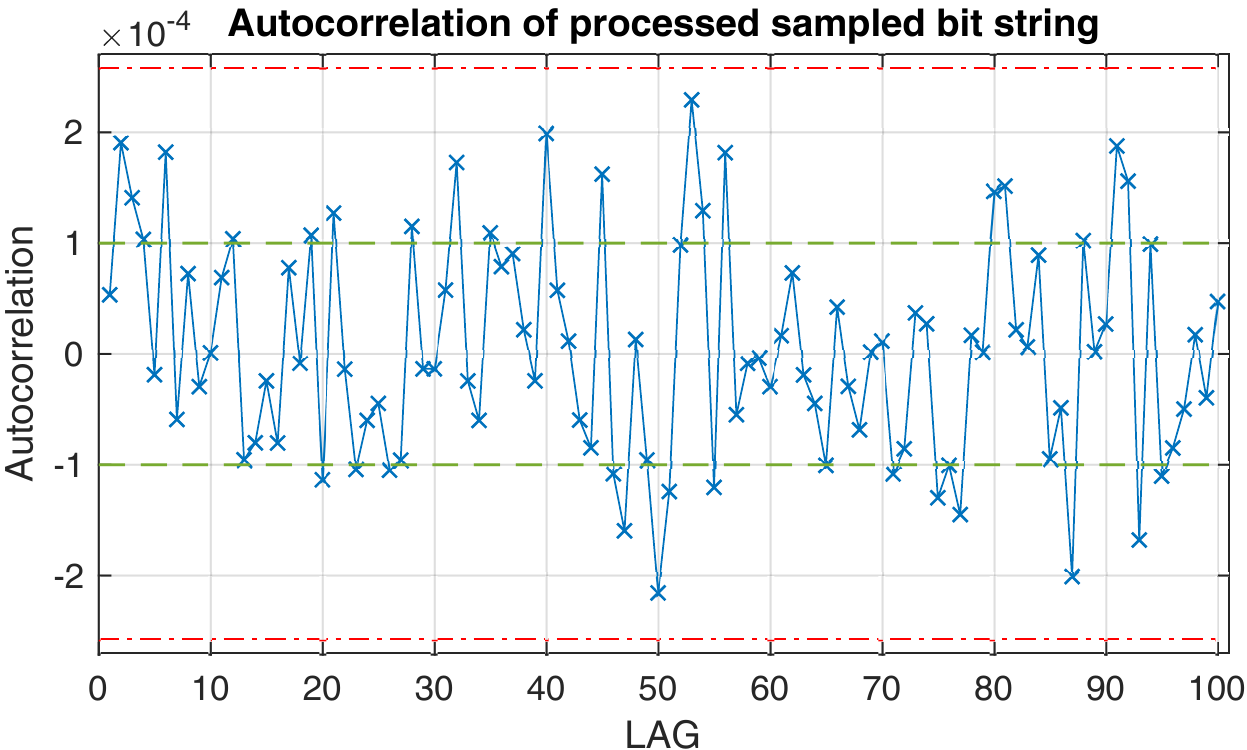}}
\caption{Top plot: the serial correlation evaluated on a sampled bit string without any dead time or after pulse treatment. Bottom plot: the serial correlation evaluated on the same sampled bit string after dead time and afterpulse removal. The serial correlation re-enters within the limit of acceptance. Green dashed lines represent standard deviation while red dot-dash lines are 99\% confidential limits.}
\label{correlationRandy}
\end{figure}

As a result, this procedure eliminated a portion of valid random events which we evaluate to be around 14\%. However, the resulting distribution of the difference between consecutive 1's follows the expected decreasing exponential (see Fig. \ref{diffRandy} bottom). Moreover, the process eliminates the correlation initially present in the string $X$, as demonstrated by Fig. \ref{correlationRandy}.

Once dead time and afterpulse are compensated, we apply Peres algorithm to the bit string.
With an actual photon count rate of 172 kcounts/s and an equivalent sampling frequency of 97 MHz (due to afterpulses and dead time removal), a real time implementation of the Peres algorithm would have yielded a final true random bit rate of 1.8 Mbit/s according to the rates shown in Fig. \ref{ratePeres}. As shown in Fig. \ref{correlationRandy} the correlation over the output string is within the limits of statistical acceptance.

Moreover, the two generation protocols Diff-QRNG and OdEven-QRNG described in \cite{Stipc2007,Furst2010} were also implemented on the same FPGA-system in order to compare their performances with
our method. The high time resolution ($r_{\text{phot}}\ll\tauclk^{-1}$) allows enough precision to identify time differences and to have a good estimation on the detected photons within a fixed intervals. Indeed, the system was successfully used to produce timed random numbers in the work of Vedovato \textit{et al.} \cite{Vedovato2017}. Of course, the only uncertainty came from the photons time of arrival but it was handled by setting the photon counting rate to a proper value \footnote{The time required by the FPGA in order to implement the protocols was totally negligible}. On the other hand, these protocols have low rate performances: a photon count rate of 200 kcount/s produce a rate of true random bits of 100 kbit/s for the Diff-QRNG protocol and of 20 kbit/s for the OdEven-QRNG \footnote{The time interval was set in order to have \unexpanded{$\langle n_\tau \rangle\simeq$} 10 according to what stated in section \ref{examples}}.

\section{Multiplexing the generation rate}\label{linospadSec}
The main limitation  concerning the use of QRNG based on SPADs, i.e. discrete variable QRNG, is the limited generation rate  achievable. 
Typically, SPADs feature maximum count rates of a few Mcps (counts per second). 
QRNGs based on continuous variable protocols are therefore preferred when it comes to obtain rates in the order of Gbps \cite{Marangon2017,Avesani2018}.
However, the recent advancements in miniaturization techniques, and especially the creation of deep-submicron CMOS SPADs, have led to arrays and matrices with hundreds or even thousands of SPADs \cite{Niclass2006,Charbon2014}.
Hence, each SPAD can be considered as a pixel of an extremely sensitive light sensor.
The application to the case of random number generation is then straightforward: given that every pixel works independently, it is possible to multiplex the random signals and then fill the rate gap with CV-QRNGs \cite{Stucki2013,Burri2013}.
The typical approach is to generalize the paradigm of the welcher weg QRNG - a beam splitter and two photon paths - to a generator where photons can take $N$ possible paths, with $N$ being the total pixel number of the sensor \cite{Marangon2016}.
Nevertheless, as in Randy, the temporal degree of freedom can be exploited as well allowing an easier calibration. Therefore, the sensor is illuminated with a uniform light intensity in order that each SPAD has the same probability to click within a given time interval, which  corresponds to the exposure time for a frame.
Random numbers are indeed produced by periodically sampling each pixel and applying dedicated generation protocols.

In this work we consider a sensor of recent introduction, \emph{LinoSPAD} \cite{Burri2016}, from the AQUA lab at Delft University \& EPFL, which features 256 pixels arranged in four linear arrays of 64 pixels each.
The peculiarity of LinoSPAD is a time tagging functionality, which associates a temporal coordinate to every detection, thus potentially enabling a further increase in the generation rate.
In the following, we will apply the techniques described in the previous Sections for the single detector case to LinoSPAD.

LinoSPAD features 64 FPGA-based time-to-digital converters (TDCs) \cite{Favi2009,Fishburn2013} that tag the detections of each SPAD in a given bank.
Each TDC is implemented by a delay line with 35 carry elements of 4 bits and it is sampled with a frequency $f_{\text{clock}}=400$ MHz. 
Every time interval $\tau_{\text{clock}}=2.5$ ns the TDC emits an output code $b \in \left\{0,139\right\}$.
The TDC has therefore a sub-resolution of $\tau_{\text{sub}}=\tau_{\text{clock}}/140 \simeq17.86$ ps and this represents the fundamental time resolution of the system. The following considerations take into account that the actual number of valid pixels is 64 and not 256 due to the limitations to 64 TDCs.
The measurements of the photon time of arrival are taken with respect to a \virgolette{reference} time signal, whose period determines the integration time of a frame. 
The buffer of the system can add output codes to a maximum of $2^{28}$ bins. 
Hence, the longest measurable time interval between the reference clock signal and a photon detection cannot be larger than $\tau_{\text{sub}} \cdot 2^{28}\simeq4.8$ ms. 

\begin{figure}[htbp]
\centering
\centerline{\includegraphics[width=\columnwidth]{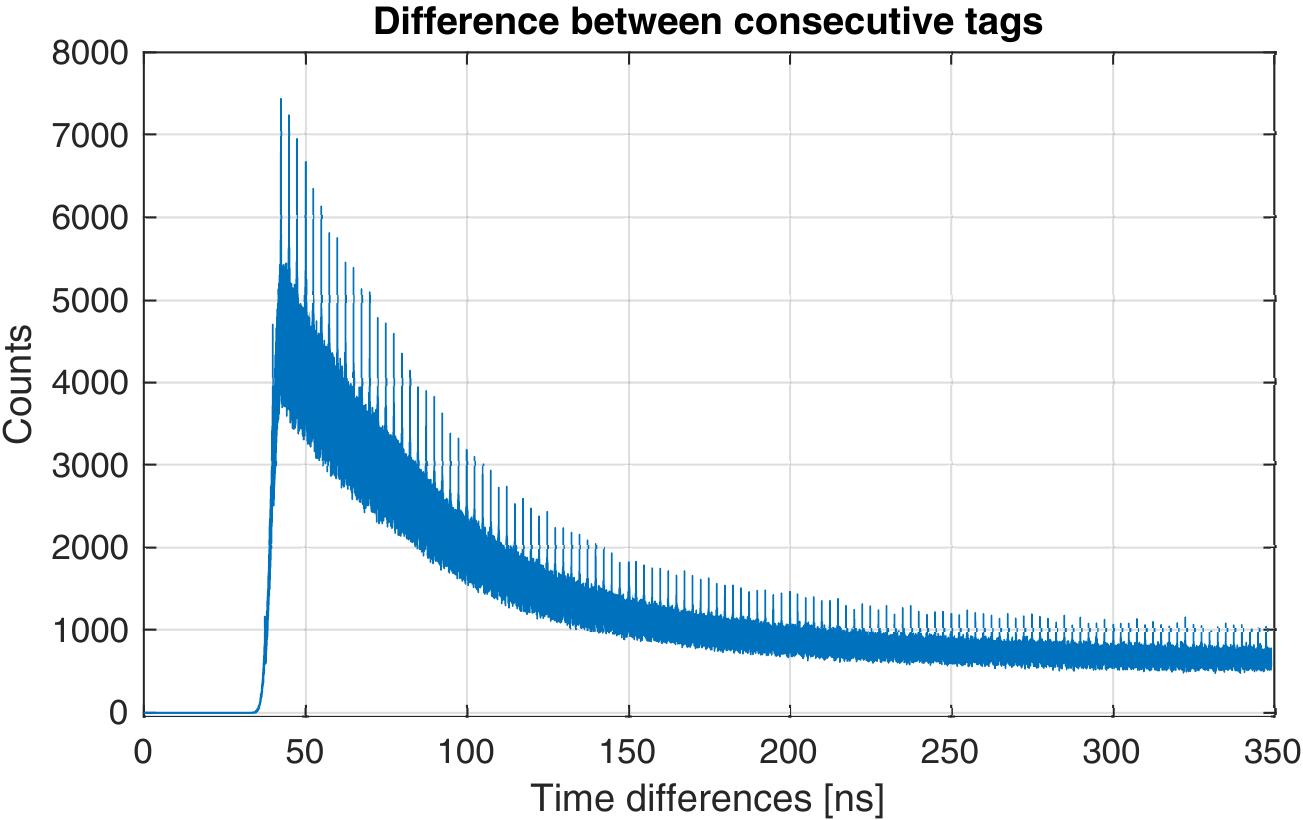}}
\caption{Distribution of the time differences between successive detections on LinoSPAD device.}
\label{Linospad1}
\end{figure}

Since the device registers a maximum of 512 tags per pixel during the integration time, every frame is composed at most by 64 x 512 tags.
Similarly to the paradigm adopted in the previous Sections, random bits can be extracted directly from the bare physics of the process: a string $\tau_{\text{frame}}/\tau_{\text{sub}}$ bits long is associated to each frame and the 1's, equal in number to the number of tags, are located according to the tag values.
Again, the limit of this approach is that the strings are consistently biased towards zero.
This bias is the result of two concurring causes: the first one is the dead time of the SPADs, which being of $\tau_{\text{dead}}\simeq$40 ns, implies that every bit 1 is necessarily followed by $\tau_{\text{dead}}/\tau_{\text{sub}}\simeq$2240 0's. 
However, the 0's due to the dead time can be removed, as it was previously done. 
The second cause is the limited buffer size: each string produced in a frame will always feature at most 512 1's.
Indeed, an extraction approach by means of the Peres debiasing procedure could be suitable even for this framework.
As in Randy, we studied the interarrival detections time, whose distribution is reported in Fig. \ref{Linospad1}, in order to detect the artefacts induced by the physical limits of the device.

The histogram starts approximately at 40 ns due to the dead time and a peak starting at 40 ns and extending up to 200 ns indicates the presence of afterpulses. A noticeable feature is an unusual peaks pattern. This pattern was due to a non-linear behaviour of the TDC (further details in subsection \ref{FineSubSec}). Hence, to manage these non-idealities we decided to separate the tag resolution between \textit{coarse} resolution (Clock sampling) and \textit{fine} resolution (TDC) implementing two different post-processing procedures. This distinction allows an easy post-processing procedure since it separates the treatment of non-idealities, i.e. \textit{coarse} for detectors ones and \textit{fine} for electronics ones (see subsection \ref{coarseSubSec} and \ref{FineSubSec}).
The analysis was done on data collected with an acquisition of $8\cdot 10^3$ frames where every frame has an integration time $\tau_{\text{frame}}=320 \cdot 10^{-6}$ s.
The photon rate was tuned to obtain approximately 400 counts per frame.
Given a buffer size of 512 detections, this value was chosen in order to keep low the probability of saturation and frame loosing.
Results show a final achieved generation bit rate equal to 310 Mbit/s.

\subsection{COARSE resolution}\label{coarseSubSec}
\begin{figure}[htbp]
\centering
\centerline{\includegraphics[width=\columnwidth]{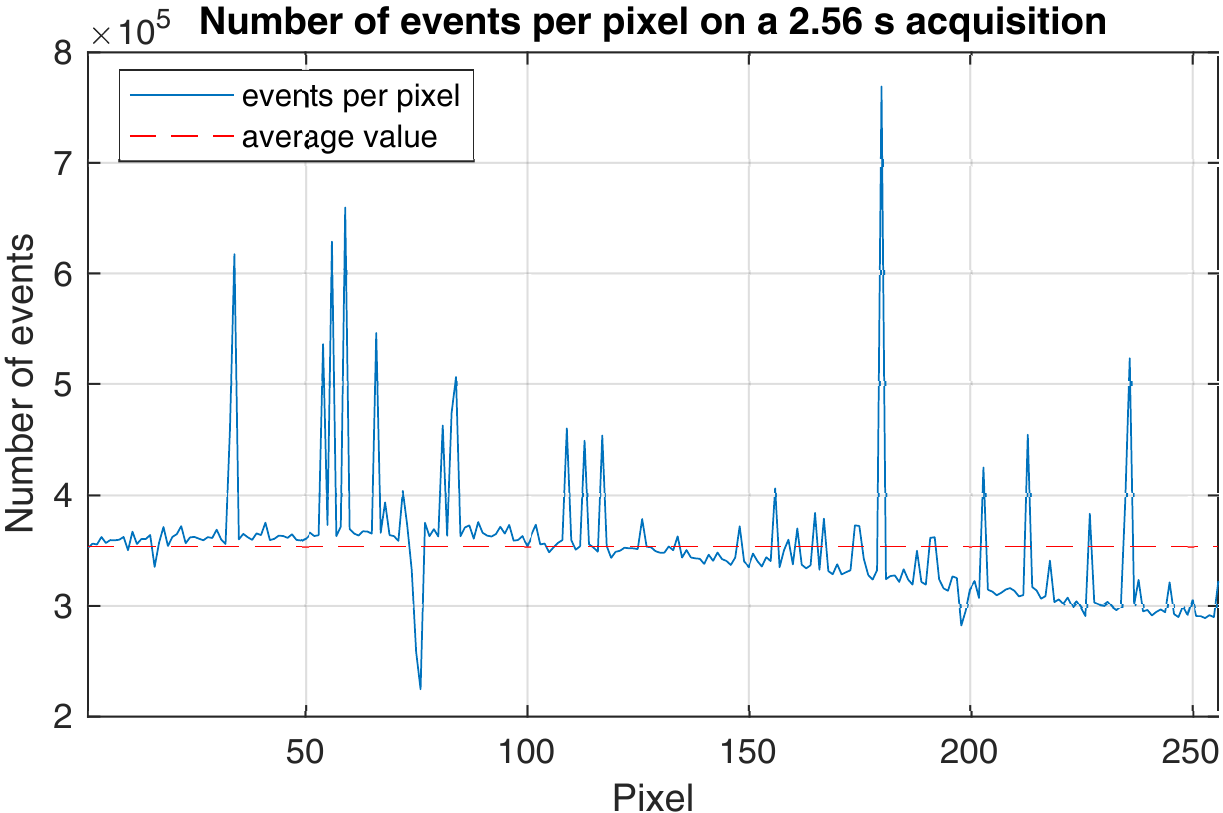}}
\caption{Number of events for every pixel. From the graph is it clear that the efficiency varies from pixel to pixel with high-efficient-isolated pixels and a descending trend.}
\label{conteggi}
\end{figure}

The \textit{coarse} resolution is defined by the 400 MHz system clock. The tag information is related to the number of clock cycles in which an event is detected. Therefore, it describes the temporal distance between an event and the zero reference with a resolution of 2.5 ns. 
We remove dead time and afterpulses of the SPADs with the same technique described in section \ref{novel}. However, the SPAD arrays suffer from \textit{pixel cross-correlation} which causes a pixel to output an event when its neighbour receives a photon, i.e. it produces a fake event as in the afterpulse. In order to remove it and to be sure that no cross correlation exists, we discard approximately one third of the pixels of a bank losing all the events from those pixels. This procedure reduces the actual number of pixel from 64 to 22. Once the cross correlation is removed as well as the dead time and the afterpulse, we apply Peres algorithm to every selected pixel independently. As discussed in section \ref{novel} and according to Fig. \ref{ratePeres}, higher rates can be achieved by preferring longer heavily biased bit strings over shorter slightly biased bit strings. Therefore, we treat every selected pixel as an autonomous QRNG.
As in the case of single detector implementation, we evaluated the serial correlation on a sampled bit string and it re-enters within the limit of acceptance after removing dead time, afterpulse and pixel cross-correlation.
By summing the bit rate of the selected pixels we obtain a total bit rate of $\mathcal{R}_{\mathrm{coarse}}\simeq87$ Mbit/s. The value is averaged since the event rate varies from pixel to pixel as shown in Fig. \ref{conteggi}. The mean extraction rate per pixel is equal to $\mathcal{R}_{\mathrm{coarse}}/64=\mathcal{R}_{\mathrm{coarse,}p}\simeq1.36$ Mbit/s. Considering an ideal SPAD array with no correlation, the hypothetical extraction rate per pixel would be equal to $\mathcal{R}_{\mathrm{coarse}}/22=\mathcal{R}_{\mathrm{coarse,}p}^{*}\simeq3.95$ Mbit/s.

\subsection{FINE resolution}\label{FineSubSec}
\begin{figure}[htp]
\centering
\centerline{\includegraphics[width=\columnwidth]{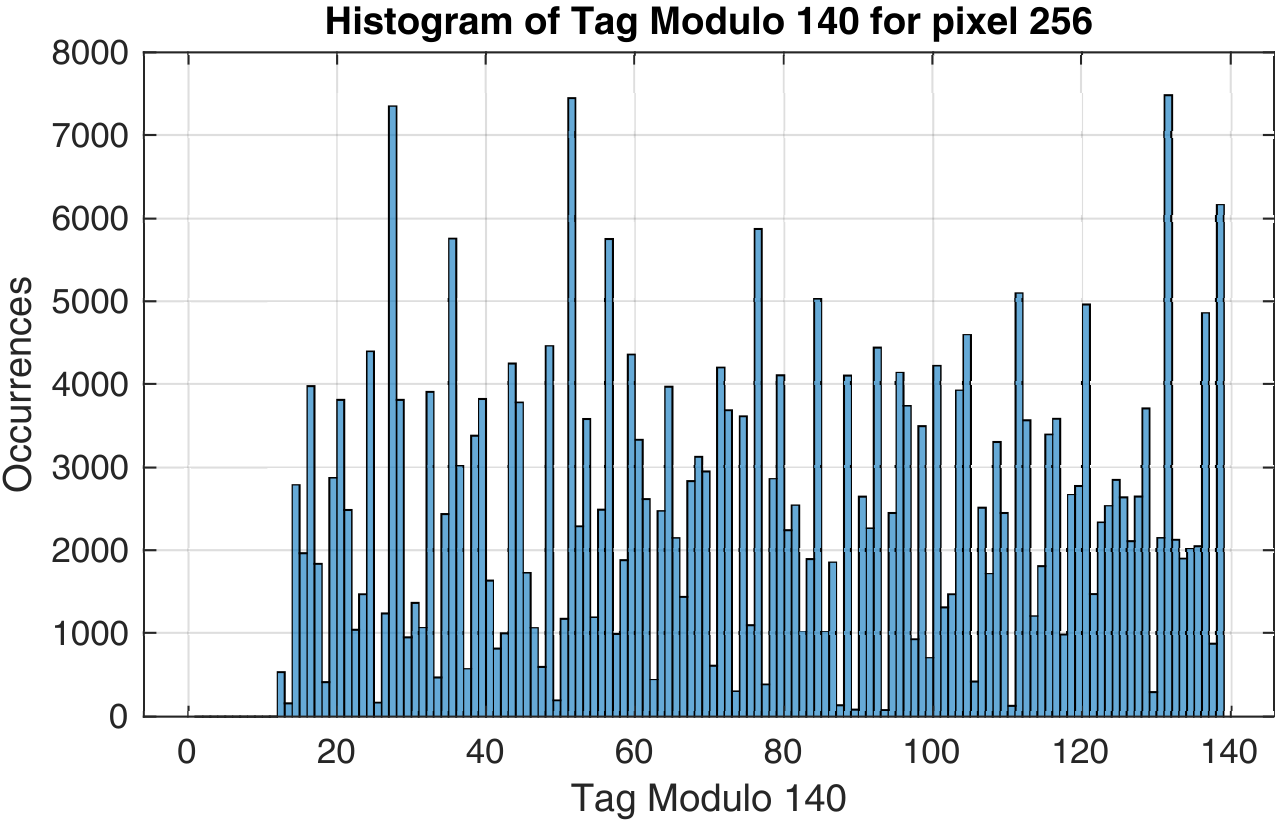}}
\caption{Distribution of the tags modulo 140 of the pixel 256. The highly biased distribution shows the non-linear response of the TDC.}
\label{TDC}
\end{figure}

The \textit{fine} resolution is defined by the TDC and has a time resolution of approximately 17.86 ps. The TDC outputs a number between 0 and 139 which identifies a precise moment within a clock cycle in which there was an event. Hence, the access to the TDC value is done every clock cycle. As in the \textit{coarse} resolution, we decide to treat every pixel independently. The peaks of Fig. \ref{Linospad1} are separated exactly by 2.5 ns, which can be explained by the presence of a dead time in accessing the TDC. This behaviour brings to the tag distribution shown in Fig. \ref{TDC} which represents an entire 2.5 ns clock cycle.
The figure shows that there are no events in the lower bins and also a significant bias on several values. This behaviour is due to the TDC implementation on an FPGA technology which introduces non-linearity caused by different propagation delays over hardware blocks \cite{Song2006}. It is worth to notice that, while the distribution is not uniform, its entropy is equal to $H_{\rm exp}=-\sum_{k=0}^{139}p_k\log_2p_k\simeq6.8$ bits, very close to the maximum entropy $H_{\rm th}=\log_2140\simeq7.12$ achievable with 140 uniformly distributed decimal values. We also point out that, if such 140 integers are transformed into their binary description, the biased distribution introduces a correlation of the bits. Numbers in decimal basis are just biased and not correlated; by transforming the integers into a binary basis, the resulting bits are correlated due to the non uniform original distribution. Moreover, in order to describe 140 different values, 8 bits are required, implying that there are no events from 140 to 255, which worsens the situation. Clearly, Peres algorithm cannot be applied to this string.

\begin{figure}[htbp]
\centering
\centerline{\includegraphics[width=\columnwidth]{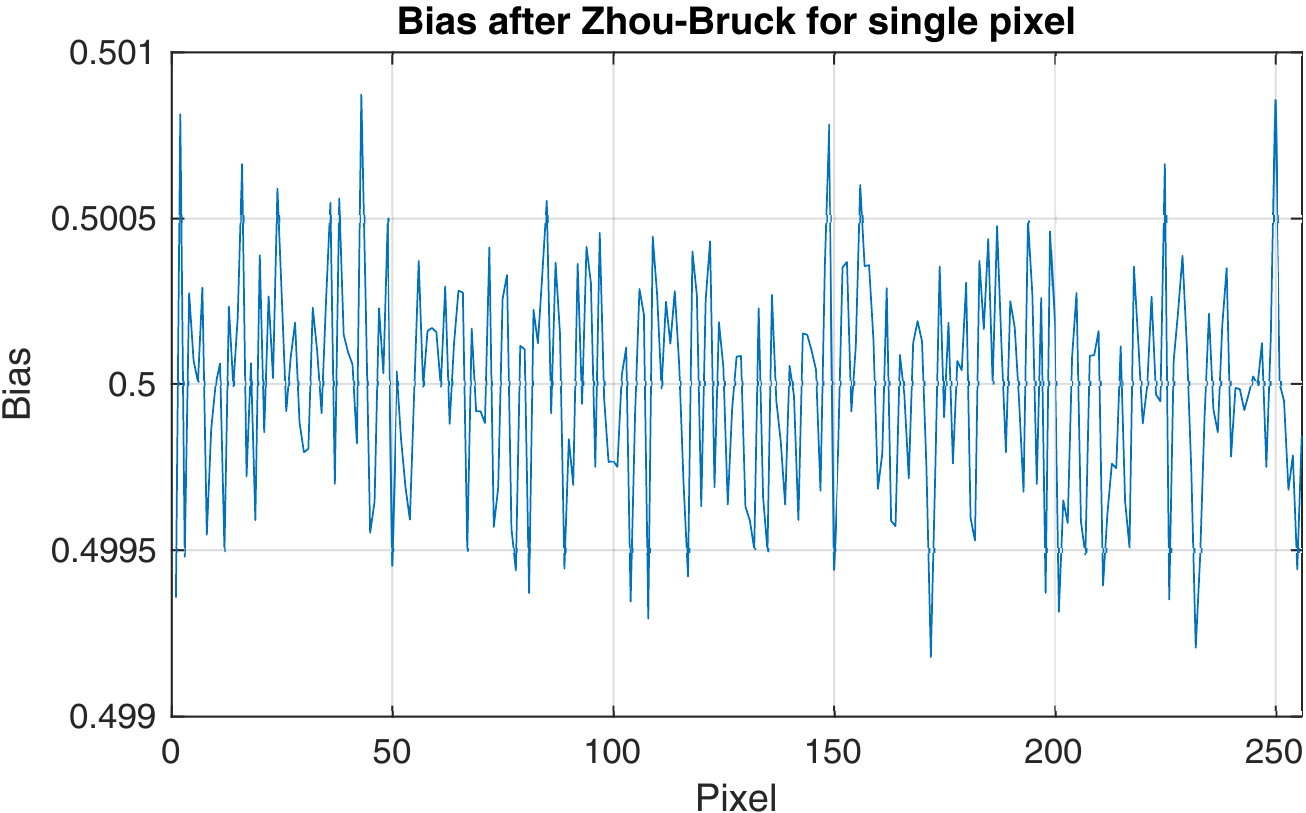}}
\caption{Bias after the application of the Zhou-Bruck Method. The graph shows the perfect balancing between 0's and 1's for every pixel within the order of $10^{-4}$.
}
\label{zhbr-bias}
\end{figure}

\begin{figure}[htbp]
\centering
\centerline{\includegraphics[width=\columnwidth]{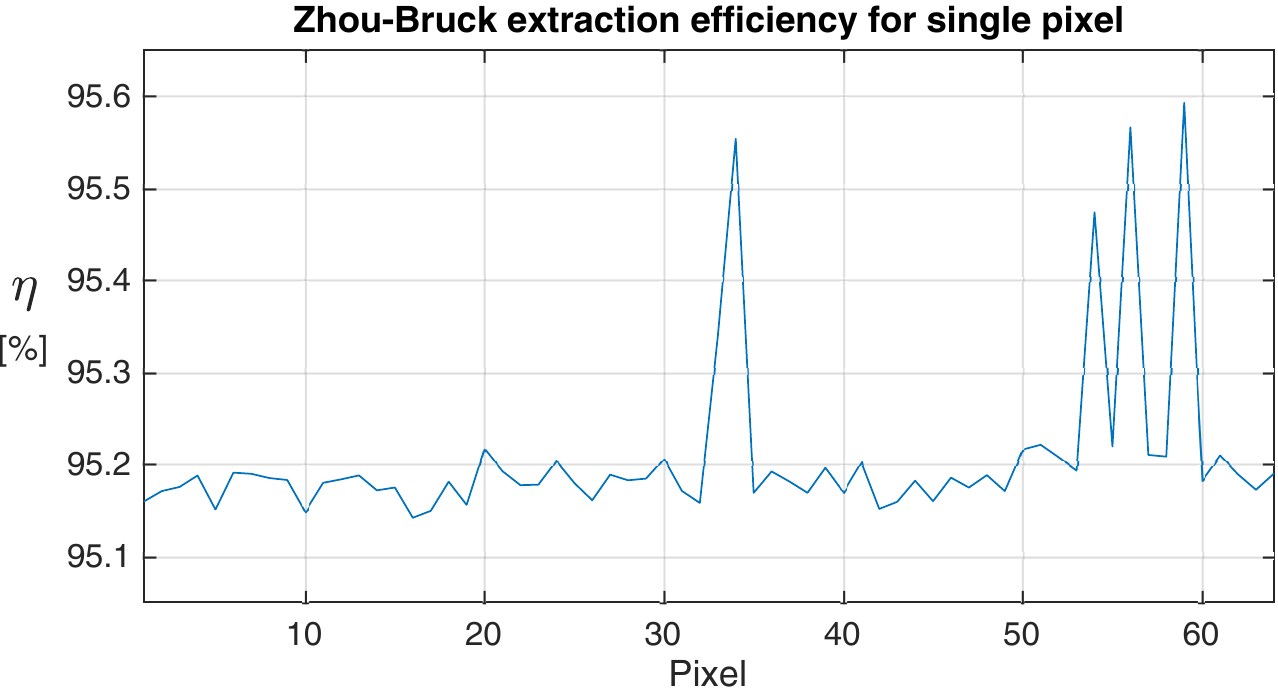}}
\caption{Extraction efficiency of the Zhou-Bruk method for every pixel. The efficiency is evaluated as $\eta(p)=N_{\mathrm{bit}}/(N_{\mathrm{tag}}\cdot H_{\rm exp})$ where $N_{\mathrm{bits}}$ and $N_{\mathrm{tags}}$ are the number of extracted bits and the number of tags. The four spikes are due to a higher count efficiency of the selected pixels according to Fig. \ref{conteggi}.}
\label{zhbr-h}
\end{figure}

In order to remove the bias and the correlation on the modulo 140 distribution, it is possible to use another post-processing method, the algorithm proposed by Zhou and Bruck~\cite{Zhou2012}. 
The latter is the generalization of the Peres algorithm for biased
distributions over a finite number of integers. Now we briefly describe the Zhou-Bruck algorithm (for a full description of the method refer to \cite{Zhou2012}).
Let's consider a random variable $X$ with $n$ possible outcomes with a biased probability distribution. Let's define 
$b=\lceil\log_2n\rceil$ as the number of bits required for a binary description of the outcomes (in our case with $n=140$ we have $b=8$).
If the outcomes are labelled as $0$, $1$, $\cdots$, $n-1$ and converted to binary, a string of $b$ bits corresponds to each outcome. 
For a given sequence $(x_1,\cdots, x_M)$, we can convert any $x_k$ to the corresponding binary string and consider only the first bit of each string: the resulting sequence is biased but with no correlation.
Let's now consider the second bit of each string. We may create two sequences corresponding to the two possible values of the first bit: the first sequence collects the second bits related to a 0 first bit and vice versa for the second one.
Again, these two sequences have bias but no correlation.
The idea can be iterated: the $i^{th}$ bits can be grouped into $2^{i}-1$ sequences according to the
values of the $i-1$ bits. 
After this manipulation, one gets $N$ different sequences where $N=2^{b}$. Since these $N$ sequences have bias but no correlation, the Peres algorithm can be applied separately to each of them (if the sequence is not empty).
We implemented the Zhou-Bruck method on the tags modulo 140 of each pixel. As a matter of fact, the Zhou-Bruck method is quite efficient: 
for example, for the pixel 256, from each tag we get an average of 6.3 unbiased bits: this value is close to the maximum $H_{\rm exp}\simeq 6.8$ that can
be extracted with a perfect efficient algorithm. The obtained bits were evaluated in term of bias (Fig. \ref{zhbr-bias}) and binary entropy extraction efficiency (Fig. \ref{zhbr-h}) as well as serial correlation which re-enters within the limit of acceptance. 

The final rate of the \textit{fine} resolution is equal to $\mathcal{R}_{\mathrm{fine}} = 223$ Mbit/s for the whole pixels array. In order to evaluate the rate for a single pixel it must be taken in account that there are only 64 TDCs which switch among the four pixel banks. Therefore, the actual number of pixels (in therms of rate) truly is 64 and the final rate per pixel is $\mathcal{R}_{\mathrm{fine}}/64=\mathcal{R}_{\mathrm{fine,}p}\simeq3.48$ Mbit/s per pixel.
Thus, by considering the \textit{fine} and \textit{coarse} resolution, the average generation rate per pixel is given by $\mathcal{R}_{\mathrm{tot,}p}=\mathcal{R}_{\mathrm{fine,}p}+\mathcal{R}_{\mathrm{coarse,}p}\simeq 4.84$ Mbit/s while the total generation rate for LinoSPAD device is given by $\mathcal{R}_{\mathrm{tot}}=\mathcal{R}_{\mathrm{fine}}+\mathcal{R}_{\mathrm{coarse}}\simeq 310$ Mbit/s.

\section{Conclusion}\label{conclusions}

\begin{figure}[tbp]
\centering
\centerline{\includegraphics[width=\columnwidth]{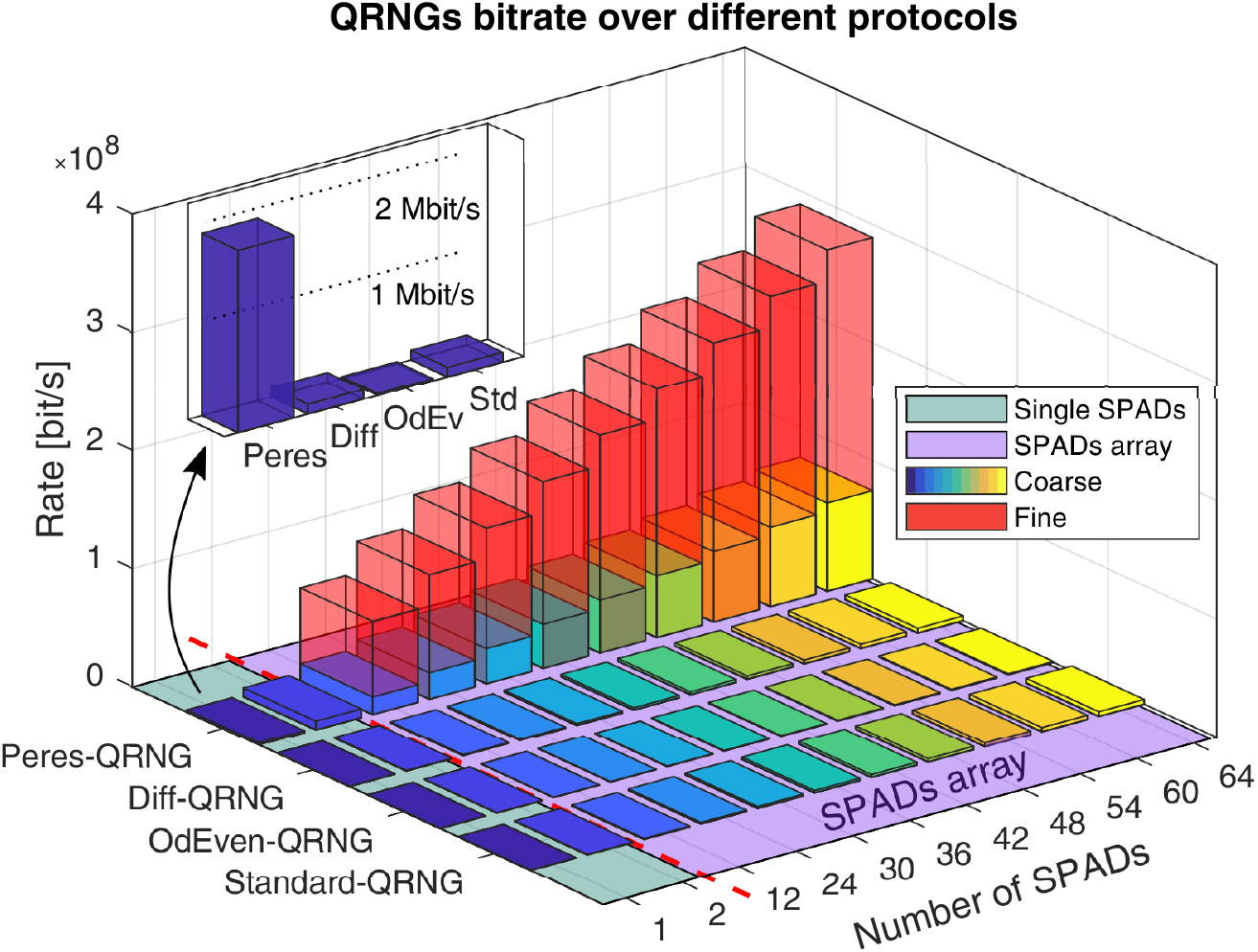}}
\caption{Bar diagram with different protocols rates comparison. The diagram shows the qualitative bit rates differences between different protocols. It is divided in two areas which represent the use of discrete SPADs (as in Randy) and SPAD array (LinoSPAD).}
\label{ratecomparison}
\end{figure}

In this paper we showed improved techniques to produce true random numbers by processing single photon events. An innovative CMOS SPAD array device called LinoSPAD was used to implement a high rate RNG. It integrates a temporal tagging system with a detector matrix and thus it allows to use the temporal degree of freedom in addition to the much more common spatial one. Starting from a single detector system (Randy), we defined an efficient procedure to fully exploit the temporal degree of freedom. With respect to existing paradigms which are based on complex rules and are far from being information efficient, our procedure extracts the most of the system entropy, achieving the maximum bit rate allowed by the system. This procedure is based on the use of a high frequency sampling clock (with respect to the photon rate) and on the use of the Peres unbiasing algorithm. Therefore, the generation rate is only limited by the physical device performances and not by the technique itself. Applying this technique to LinoSPAD required a further step to deal with detectors matrix non-idealities (pixel cross correlation) and TDC ones. Hence, a dedicated post-processing procedure, which included the usage of the Zhou-Bruk algorithm, was developed in order to work around such non-idealities. The summary comparison bar diagram of Fig. \ref{ratecomparison} clearly shows the remarkable differences among different generation procedures. Our Peres-based procedure clearly reaches a higher bit rate with respect to a protocol-based one. Furthermore, moving from the discrete SPADs framework (Randy) to the CMOS SPAD array one and increasing the sampling frequency as well as adding a time-to-digital converter (LinoSPAD) improves the generation rate even more. Final results show a bitrate per SPAD/pixel equal to:
\begin{itemize}
    \item $\mathcal{R}_{\mathrm{Randy,}spad} (100\: \mathrm{MHz}) = 1.8$ Mbit/s
    \item $\mathcal{R}_{\mathrm{LinoSPAD,}spad} (400\: \mathrm{MHz} + \mathrm{TDC}) = 4.84$ Mbit/s
\end{itemize}
and for LinoSPAD a total bitrate of
$\mathcal{R}_{\mathrm{LinoSPAD}} = \mathcal{R}_{\mathrm{LinoSPAD,}spad} \times 64 = 310$ Mbit/s.

Moreover, applying these techniques to other physical devices with better performances will further increase  the generation rate. Future steps will also consider a real-time implementation of both procedures since the pre-processing, the Peres and Zhou-Bruk algorithms could be effectively implemented via FPGA.

\bibliographystyle{apsrev4-1}
\bibliography{bib_resource}

\end{document}